\documentclass[%
 reprint,
 amsmath,amssymb,
 aps,
floatfix,
]{revtex4-2}

\usepackage{multirow}

\usepackage{graphicx}
\usepackage{dcolumn}
\usepackage{bm}
\usepackage{hyperref}
\hypersetup{
    colorlinks = true,
    linkcolor = blue,
    anchorcolor = blue,
    citecolor = blue,
    filecolor = blue,
    urlcolor = blue
    }
\usepackage{amsmath}
\usepackage{orcidlink}

\begin{document}

\preprint{APS/123-QED}

\title{The study on the multiplicity dependence of ridge behavior in $pp$ collisions at $\sqrt{\textbf{s}}=13$ TeV at the LHC}

\author{Jeongseok Yoon\orcidlink{0009-0002-7208-0903}}
 \email{jeongseok.yoon@cern.ch}
\author{Jin-Hee Yoon\orcidlink{0000-0001-7676-0821}}%
 \email{jinyoon@inha.ac.kr}
\affiliation{%
 Department of Physics, Inha University, Incheon 22212, Republic of Korea
}%




\date{\today}

\begin{abstract}
The long-range near-side ridge phenomenon in two-particle correlation is crucial for understanding the motion of partons after high-energy heavy-ion collisions. While it has been well explained by the hydrodynamic flow effect of the quark-gluon plasma (QGP) in heavy-ion collisions, the recent observation of the ridge structure in small systems has led to debates about the applicability of hydrodynamic models to explain the phenomenon since the collisions in small systems could not be sufficient to produce the medium required by the QGP matter. The Momentum Kick Model (MKM), on the other hand, explains the long-range near-side ridge phenomenon by the kinematic process; the high-momentum jet particles collide with medium partons, transfer their momentum to them (called the ``kick" process), and induce collective motion of the kicked-partons resulting in the ridge phenomenon. This MKM has successfully described the ridge structure in heavy-ion collisions at the RHIC. Furthermore, since the ridge phenomenon in small systems is prominent in high-multiplicity events, the MKM with multiplicity dependence (MKMwM) has been studied in $pp$ collisions at the LHC using a relationship between the number of kicked-partons and the multiplicity through an impact parameter. In this research, we extend the previous study with more recent experimental data-driven parameters and apply them to the new measurements that have a wider multiplicity range with $p_T$ and $\Delta\Phi$ bins at the LHC. Simultaneously, we not only provide a theoretical basis for the ridge behavior from the new measurements but also predict the ridge structure at the energies scheduled by the LHC in the upcoming Run 3 experiments.
\end{abstract}

\maketitle


\section{INTRODUCTION}\label{sec:INTRODUCTION}

High-energy heavy-ion collision experiments have provided valuable insight into the nature of fundamental particles. Among them, the two-particle correlation between the trigger and the associated particles has revealed many interesting physical phenomena. Naively, the long-range near-side ridge phenomenon emerges in this two-particle correlation and has a significant impact on our understanding of the interacting process in heavy-ion collisions. In particular, there is a long and storied history of the ridge phenomenon, which was initially observed in heavy-ion collisions, such as AuAu collisions at the RHIC~\cite{exp:STAR_AuAu_200GeV_1,exp:STAR_AuAu_200GeV_3,exp:STAR_AuAu_200GeV_5,exp:STAR_AuAu_200GeV_6e,exp:STAR_AuAu_200GeV_11,exp:STAR_AuAu_200GeV_2,exp:STAR_AuAu_200GeV_4,exp:STAR_AuAu_200GeV_6,exp:STAR_AuAu_200GeV_7,exp:STAR_AuAu_200GeV_8,exp:STAR_AuAu_200GeV_9,exp:STAR_AuAu_200GeV_10,exp:STAR_AuAu_200GeV_12,exp:PHENIX_AuAu_200GeV_0,exp:PHENIX_AuAu_200GeV_1,exp:PHENIX_AuAu_200GeV_2,exp:PHENIX_AuAu_200GeV_3,exp:PHENIX_AuAu_200GeV_4,exp:PHOBOS_AuAu_200GeV_1} and PbPb collisions at the LHC~\cite{exp:CMS_PbPb_276TeV_1,exp:CMS_PbPb_276TeV_2,exp:CMS_PbPb_276TeV_3,exp:CMS_PbPb_276TeV_5,exp:CMS_PbPb_276TeV_4,exp:CMS_PbPb_276TeV_6,exp:CMS_PbPb_276TeV_7,exp:ALICE_PbPb_276TeV}. Most physicists have believed that the hydrodynamic flow effect~\cite{hydro1,hydro2,hydro3,hydro4,hydro5,hydro6} of the quark-gluon plasma (QGP) causes ridge behavior in a high-temperature and high-density medium. However, as high-energy collision experiments have become possible with the LHC era, the ridge phenomenon has also been measured even in small systems, such as $pp$ and $p$Pb collisions, at the LHC~\cite{exp:CMS_pp_7TeV,exp:CMS_pp_13TeV,exp:CMS_pp_13TeV_2,exp:ALICE_pp_13TeV,exp:ATLAS_pp_276TeV,exp:ALICE_pPb_502TeV_1,exp:ALICE_pPb_502TeV_2} in high-multiplicity events. This has raised questions about the applicability of previous hydrodynamic models to explain the ridge phenomenon in small systems that are not believed enough to generate the QGP medium.

Consequently, several models have been proposed to account for the ridge phenomenon in small systems. For example, the correlated emission model~\cite{corr_emis_model} assumes that successive soft emissions from the jet-medium interaction lead to increasing thermal partons. Furthermore, a multiphase transport (AMPT) model~\cite{AMPT_model1,AMPT_model2} explains initial partonic and final partonic interactions using the heavy ion jet interaction generator (HIJING). In terms of multiple parton interactions (MPIs), the color reconnection (CR)~\cite{CR1,CR2} and a hydrodynamic mechanism~\cite{MPIs1,MPIs2} are introduced. In particular, the eikonal model~\cite{Eikonel} for MPIs includes the collective expansion effects in $pp$ collisions at the LHC. The gluon saturation model~\cite{gluon_saturation_model}, on the other hand, is founded on the premise that multiparticle production is controlled by a singular semi-hard saturation scale within the color glass condensate framework. However, a conclusive explanation for the ridge phenomenon in small systems has not yet been established.

In this paper, the Momentum Kick Model (MKM) offers an explanatory framework for the ridge phenomenon through a simple kinematic process~\cite{Taewook,Beomkyu,Wong:STAR_AuAu_200GeV,Wong:PHENIX_AuAu_200GeV,Wong:CMS_pp_7TeV,Wong:Initial_Form,Wong:extra_explan}. Jet particles with relatively high momentum pass through the medium, collide with nearby partons, and transfer their momentum to medium partons in a process called a ``kick''. These kicked-partons move collectively along with the jet particles, where this collective motion becomes the ridge phenomenon. In particular, since the backward ridge is complicated due to the multiple scattering effects, the MKM focuses on the ridge behavior across the long-range ($\Delta\eta>2$) at the near-side ($\Delta\Phi\approx0$) in two-particle correlation. The kinematic process of the MKM well-described experimental data~\cite{Wong:STAR_AuAu_200GeV,Wong:PHENIX_AuAu_200GeV,Taewook,Beomkyu}; from the STAR~\cite{exp:STAR_AuAu_200GeV_1,exp:STAR_AuAu_200GeV_3,exp:STAR_AuAu_200GeV_5,exp:STAR_AuAu_200GeV_6e,exp:STAR_AuAu_200GeV_11}, the PHENIX~\cite{exp:PHENIX_AuAu_200GeV_0,exp:PHENIX_AuAu_200GeV_1}, and the PHOBOS~\cite{exp:PHOBOS_AuAu_200GeV_1} for AuAu collisions at $\sqrt{s_\mathrm{NN}}=0.2\;\mathrm{TeV}$; from the CMS~\cite{exp:CMS_PbPb_276TeV_1} for PbPb collisions at $\sqrt{s_\mathrm{NN}}=2.76\;\mathrm{TeV}$; from the CMS~\cite{exp:CMS_pp_13TeV} and the ATLAS~\cite{exp:ATLAS_pp_276TeV} for $pp$ collisions at $\sqrt{s}=13$ TeV. On the other hand, the ridge behavior observed in small systems is different with multiplicity, especially prominent in high-multiplicity events. However, there is no theoretical consensus on how the ridge yield changes according to the multiplicity. For this reason, the previous study~\cite{Wong:STAR_AuAu_200GeV,Wong:CMS_pp_7TeV} developed the MKM with multiplicity dependence (MKMwM) by linking the number of kicked-partons with the multiplicity through an impact parameter and applied this model to the CMS experimental data for $pp$ collisions at $\sqrt{s}=7\;\mathrm{TeV}$~\cite{exp:CMS_pp_7TeV}. Although the previous results show good agreement, some trends deviate from the experimental uncertainties. Furthermore, the experimental data at $\sqrt{s}=7$ TeV are not sufficient to clearly identify the multiplicity-dependent ridge phenomenon due to the limited number of associated yield bins in the multiplicity range. In parallel, it remains unclear whether the multiplicity dependence of the ridge phenomenon varies with the collision energy. Therefore, it is essential to verify this MKMwM over a wide range of multiplicities and collision energies to better understand the ridge phenomenon according to multiplicity in small systems.

Recently, the CMS Collaboration observed the long-range near-side ridge structure for $pp$ collisions at $\sqrt{s}=13\;\mathrm{TeV}$~\cite{exp:CMS_pp_13TeV}, which newly includes the correlated azimuthal angular ($\Delta\Phi$) distribution categorized along the transverse momentum ($p_T$) and the charged particle multiplicity ($N_\mathrm{ch}$) ranges. They also provided the integrated yields vs. $p_T$ in high-multiplicity events, and those vs. $N_\mathrm{ch}$ in the middle $p_T$ regions. From these results, the CMS Collaboration raised two questions:
\begin{enumerate}
    \item The ridge yield reaches a maximum around $p_T\approx1$ GeV/c and decreases with increasing $p_T$.
    \item The ridge structure shows an approximately linear increase with $N_\mathrm{ch}$.
\end{enumerate}
The MKMwM might give a theoretical basis for these two questions. Furthermore, the CMS Collaboration compared the ridge yield for $pp$ collisions at $\sqrt{s}=7$ TeV with that at $\sqrt{s}=13$ TeV and suggested that there is no clear collision energy dependence for the ridge structure in $pp$ collisions. This suggestion allows us to predict the ridge yield for $pp$ collisions at other collision energies, such as $\sqrt{s}=5.3$ and $8.5$ TeV, scheduled at the upcoming LHC Run 3.

The previous study~\cite{Wong:CMS_pp_7TeV} at $\sqrt{s}=7$ TeV focused on an integrated ridge yield over $\Delta\Phi$; the $N_\mathrm{ch}$ distribution of the integrated yield four $p_T$ ranges. On the other hand, our study at $\sqrt{s}=13$ TeV shifts its focus to a detailed analysis of the ridge behavior using individual yields, not integrated; the $\Delta\Phi$ distribution for both four $p_T$ and four $N_\mathrm{ch}$ ranges. We also introduce physical extensions and modifications using new experimental findings to improve the MKMwM. Furthermore, the experimental data at $\sqrt{s}=13$ TeV compared to $\sqrt{s}=7$ TeV not only extend the upper limit of $N_\mathrm{ch}$ range from 120 to 180 but also increase the number of data points by about four times, allowing us to check the effectiveness of the model more precisely.

First, we provide a qualitative description of the MKMwM in Sec.~\ref{sec:MOMENTUM KICK MODEL WITH MULTIPLICITY DEPENDENCE}, which consists of two subsections: the MKM evaluation in Subsec.~\ref{subsec:MOMENTUM KICK MODEL} and the multiplicity dependence formalism in Subsec.~\ref{subsec:MULTIPLICITY DEPENDENCE}. We then present the application of the MKMwM to the new CMS data and the prediction of the upcoming LHC Run 3 in Sec.~\ref{sec:ANALYSIS}. Finally, we draw conclusions and discussions in Sec.~\ref{sec:CONCLUSIONS AND DISCUSSIONS}.

\section{MOMENTUM KICK MODEL WITH MULTIPLICITY DEPENDENCE}\label{sec:MOMENTUM KICK MODEL WITH MULTIPLICITY DEPENDENCE}
The MKMwM is specifically elaborated in Refs.~\cite{Wong:STAR_AuAu_200GeV,Wong:PHENIX_AuAu_200GeV,Wong:CMS_pp_7TeV}. Therefore, in this section, we simply illustrate the formalism of the MKMwM.

\subsection{MOMENTUM KICK MODEL}\label{subsec:MOMENTUM KICK MODEL}
\begin{figure}[b]
    \centering\includegraphics[width=0.3\textwidth]{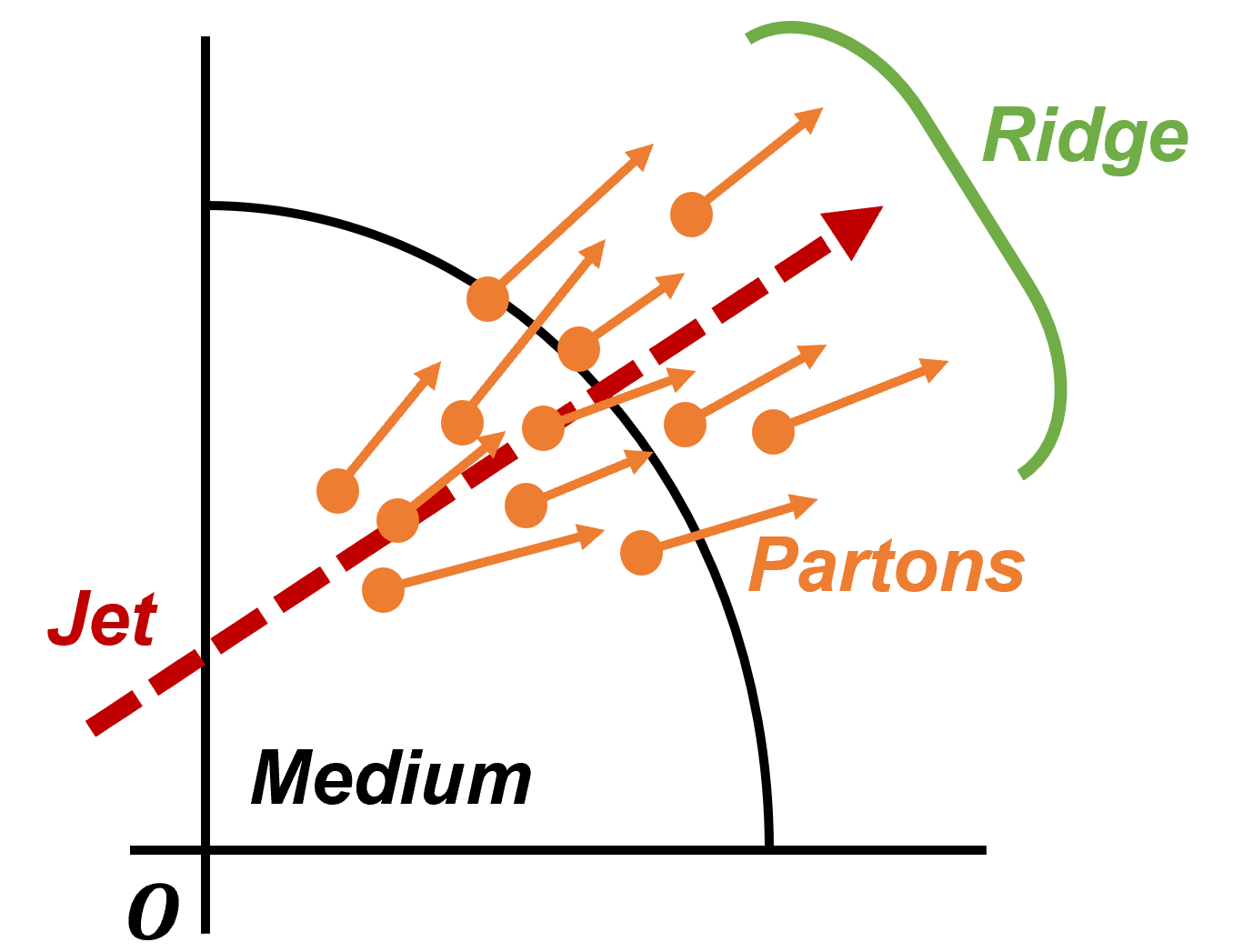}
    \caption{\label{fig:Ridge Mechanism}(Color online) Illustration for the collective motion of the kicked-partons alongside a jet particle in the transverse coordinate system, demonstrating the ridge phenomenon in the MKM. The red dashed arrow represents a jet particle, and the orange circles depict the medium partons, with the arrows indicating the momentum transfer by the jet particle.}
\end{figure}
The ridge phenomenon is a broad distribution across the long-range ($|\Delta\eta|>2$) at the near-side ($\Delta\Phi\approx0$) in two-particle correlation. The correlation function is experimentally determined as a function of the correlated pseudorapidity ($\Delta\eta$) and the correlated azimuthal angle ($\Delta\Phi$) between the trigger and the associated particles, such as $\Delta\eta = \eta_\mathrm{assoc}-\eta_\mathrm{trig}$ and $\Delta\Phi = \phi_\mathrm{assoc}-\phi_\mathrm{trig}$. In the MKM framework, the collective motion of the kicked-partons is induced by the jet particles, as depicted in Fig.~\ref{fig:Ridge Mechanism}. Therefore, the trigger and the associated particles can be regarded as the jet particles and the kicked-partons, respectively, such as $\Delta\Phi = \phi_f-\phi_\mathrm{jet}$ and $\Delta\eta = \eta_f-\eta_\mathrm{jet}$.

In addition, the motion of the medium partons is determined from the initial parton momentum distribution, which is not directly accessible because we only have information on the final state. To overcome this problem, the previous studies~\cite{Wong:Initial_Form,Wong:STAR_AuAu_200GeV,Wong:PHENIX_AuAu_200GeV,Wong:CMS_pp_7TeV,Wong:extra_explan} constructed a phenomenological initial parton momentum distribution based on experimental data from the STAR~\cite{exp:STAR_AuAu_200GeV_1,exp:STAR_AuAu_200GeV_3,exp:STAR_AuAu_200GeV_5,exp:STAR_AuAu_200GeV_6e,exp:STAR_AuAu_200GeV_11}, the PHENIX~\cite{exp:PHENIX_AuAu_200GeV_0,exp:PHENIX_AuAu_200GeV_1} and the PHOBOS~\cite{exp:PHOBOS_AuAu_200GeV_1} using a soft scattering model:
\begin{equation}\label{eq:initial parton momentum distribution}
    \frac{dF}{p_{T_i} dp_{T_i} dy_i d\phi_i} = A_\mathrm{ridge}(1-x)^a \frac{e^{-m_{T_i}/T}}{\sqrt{m_d^2+p_{T_i}^2}},
\end{equation}
where $T$, $m_\pi$, $m_d$, and $A_\mathrm{ridge}$ represent medium temperature, pion mass, mass parameter, and a normalization constant, respectively. Additionally, $a$ is the fall-off parameter that determines the falling shape of the parton distribution function along the $\eta$-direction, and the transverse mass is defined by $m_T=\sqrt{m^2+p^2_T}$, where the medium partons in heavy-ion collisions are dominated by pion, so the $m$ is set to $m_\pi$. In Eq.~\eqref{eq:initial parton momentum distribution}, $x$ is a light-cone variable given by
\begin{equation}
    x = \frac{p_{i0}+p_{i3}}{p_{b0}+p_{b3}} = \frac{\sqrt{m_\pi^2+p_{T_i}^2}}{m_b} e^{|y_i|-y_b},
\end{equation}
where the beam rapidity is denoted by $y_b=\mathrm{cosh}^{-1}(\sqrt{s_{NN}}/2m_b)$, and the beam mass $m_b$ corresponds to the mass of a proton. Since the amplitude of the forward light-cone momentum of the medium partons $(p_{i0}+p_{i3})$ cannot exceed that of the beam particles $(p_{b0}+p_{b3})$, we have the kinematic boundary condition $x\le1$.

The average momentum transfer ($\textbf{\textit{q}}$) from the jet particles to the medium partons converts the initial partons into the kicked-partons, resulting in the final parton momentum distribution. This process can be expressed by $\textbf{\textit{p}}_f=\textbf{\textit{p}}_i+\textbf{\textit{q}}$, where the lower indices, $i$ and $f$, mean ``initial states" and ``final states", respectively. In our analysis, we assume the pseudorapidity of the jet particles to be zero, considering that the majority of the jet particles lie in the transverse plane. Consequently, the normalized final parton momentum distribution can be obtained through Eq.~\eqref{eq:initial parton momentum distribution} as
\begin{eqnarray}\label{eq:final parton momentum distribution}
    \frac{dF}{p_{T_f} dp_{T_f} d\eta_f d\phi_f} &=& \left[ \frac{dF}{p_{T_i} dp_{T_i} dy_i d\phi_i} \frac{E_f}{E_i} \right]_{\textbf{\textit{p}}_i=\textbf{\textit{p}}_f-\textbf{\textit{q}}} \nonumber\\
    &&\times \sqrt{1-\frac{m_\pi^2}{(m_\pi^2+p_{T_f}^2)\cosh^2{y_f}}},
\end{eqnarray}
where the $E_f/E_i$ is the factor for the Lorentz invariance, and the last factor corresponds to the Jacobian that converts rapidity ($y$) to pseudorapidity ($\eta$).

Finally, the total ridge yield per trigger is formulated as
\begin{eqnarray}\label{eq:The total associated yield per trigger}
    &&\left[ \frac{1}{N_\mathrm{trig}}\frac{dN_\mathrm{ch}}{p_{T_f} dp_{T_f} d\Delta \eta d\Delta \phi} \right]_\mathrm{ridge} \nonumber\\
    &&= f_R \frac{2}{3} \langle N_k \rangle \left[ \frac{dF}{p_{T_f} dp_{T_f} d\Delta \eta d\Delta \phi} \right],
\end{eqnarray}
where $f_R$ plays the physical role of the survival factor for the ridge particles to reach the detector. However, it also reflects the degree of freedom of the experimental noise reduction. The factor $2/3$ corresponds to the ratio of the charged particles among the total particles, and $\langle N_k\rangle$ denotes the average number of kicked-partons.

\subsection{MULTIPLICITY DEPENDENCE}\label{subsec:MULTIPLICITY DEPENDENCE}
The ridge structure in high-energy experiments for $pp$ collisions has been extensively observed~\cite{exp:CMS_pp_7TeV,exp:CMS_pp_13TeV,exp:CMS_pp_13TeV_2,exp:ALICE_pp_13TeV,exp:ATLAS_pp_276TeV}. In particular, the ridge yield exhibits a notable increase with the charged particle multiplicity ($N_\mathrm{ch}$), especially leading to clear ridge behavior in high-multiplicity events. It is therefore plausible to think that the increasing number of charged particles increases the parton density, which naturally leads to the growing average number of kicked-partons ($\langle N_k \rangle$) in the MKM. Morevoer, since $N_\mathrm{ch}$ is inherently dependent on the impact parameter ($\textbf{\textit{b}}$), we establish a relation between $\langle N_k \rangle$ and $N_\mathrm{ch}$ through $\textbf{\textit{b}}$.

\begin{figure}[t]
    \centering\includegraphics[width=0.45\textwidth]{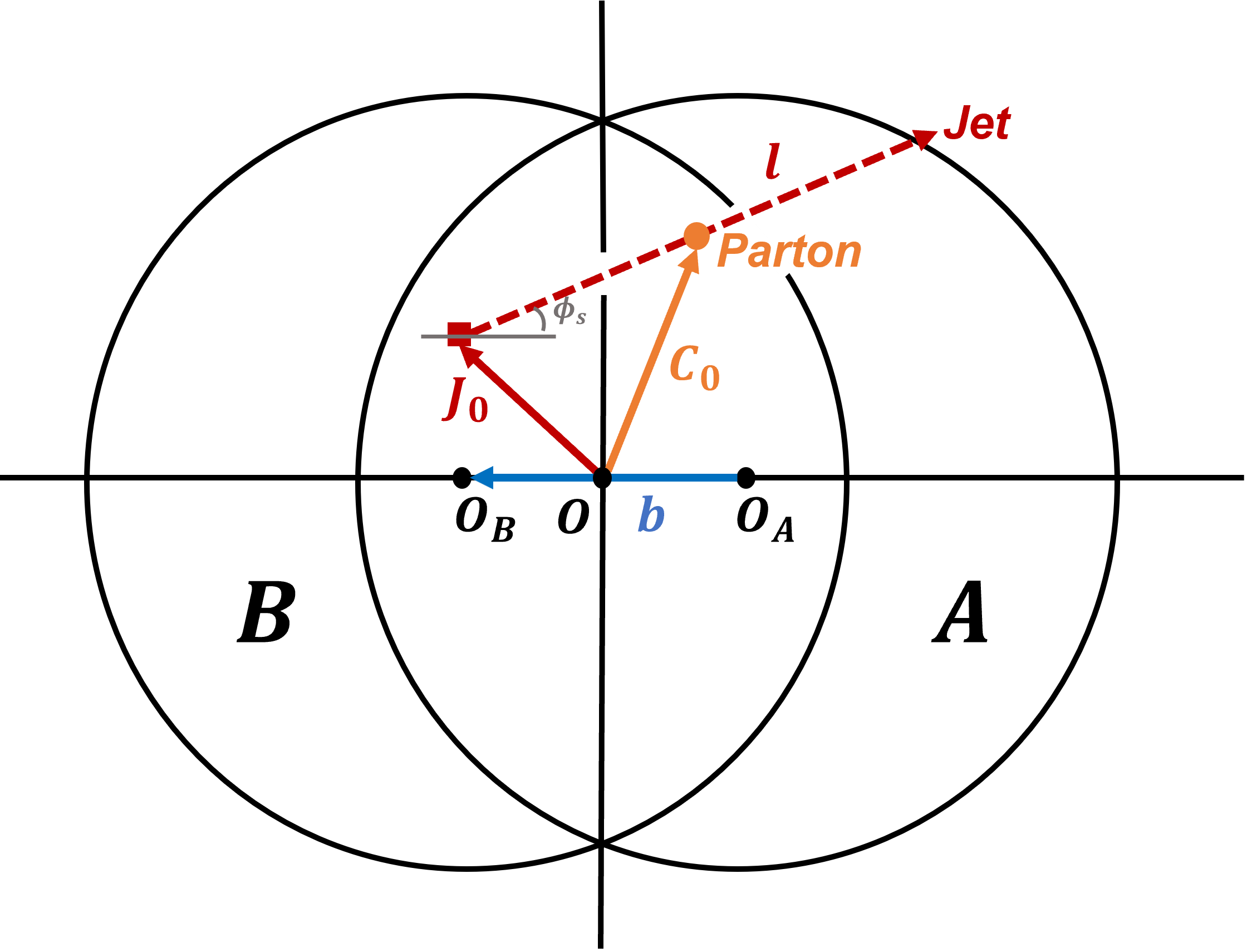}
    \caption{\label{fig:Geometry}(Color online) A schematic representation of particles $A$ and $B$ colliding with an impact parameter $\textbf{\textit{b}}$ in a transverse coordinate system. A jet particle is depicted by a red dashed arrow, and a medium parton is represented by an orange circle. After the collision, the jet particle is generated at $\textbf{\textit{J}}_0$, and then traverses the medium along the trajectory vector $\textbf{\textit{l}}$ at an angle $\phi_s$ in the reaction plane, colliding with the medium parton at $\textbf{\textit{C}}_0$.}
\end{figure}

If we look at the collisions between $A$ and $B$ particles in the transverse section relative to the beam direction, as shown in Fig.~\ref{fig:Geometry}, a medium is formed in the overlapping region of the two particles, and the time $t$ starts clocking when the overlapping region reaches its maximum. At that moment, the initial momentum of the beam particle heading toward the longitudinal direction gradually converts to the transverse direction. When the transverse momentum becomes large enough to produce particles at time $t_0$, the jet particles can be generated at time $t$ ($t\ge t_0$) and position $\textbf{\textit{J}}_0$. The production probability of the jet particles is given by multiplying the thickness function of the colliding particles:
\begin{eqnarray}\label{eq:Jet production probability}
    P_\mathrm{jet}(\textbf{\textit{J}}_0,\textbf{\textit{b}})=N(\textbf{\textit{b}})\,T_A(\textbf{\textit{J}}_0)\,T_B(\textbf{\textit{J}}_0),
\end{eqnarray}
where $N(\textbf{\textit{b}})$ is the normalization function with respect to $\textbf{\textit{b}}$. We can find the thickness function for the proton from its charge density~\cite{ProtonDensity}, which is given by
\begin{equation}\label{eq:thickness function}
    T(\textbf{\textit{s}},\textbf{\textit{b}})=N\,\frac{4.7}{1.3+\exp{(5.1\cdot\textbf{\textit{s}}})}\,\Theta(R-\textbf{\textit{b}}),
\end{equation}
where $N$ is a normalization constant, $R$ is the radius of the colliding particles, $\textbf{\textit{s}}$ denotes a position vector from the center of the particle, and the step function $\Theta(R-\textbf{\textit{b}})$ confines the production of the jet particles within the overlap region. Whereas, the number of medium partons is proportional to the number of participants, which is given by the sum of the thickness functions for colliding particles, such as
\begin{equation}\label{eq:NMP}
    N_\mathrm{MP}(\textbf{\textit{s}},\textbf{\textit{b}}) = \kappa'\big[T_A(\textbf{\textit{s}},\textbf{\textit{b}})+T_B(\textbf{\textit{s}},\textbf{\textit{b}})\big],
\end{equation}
where
\begin{equation}\label{eq:kappa'}
    \kappa'=\frac{\langle N_\mathrm{MP} \rangle}{\langle N_\mathrm{particp} \rangle}.
\end{equation}
Therefore, $N_\mathrm{ch}$ as a function of $\textbf{\textit{b}}$ is obtained by integrating the Eq.\eqref{eq:NMP} over $\textbf{\textit{s}}$:
\begin{equation}\label{eq:Nch}
    N_\mathrm{ch}(\textbf{\textit{b}}) = \frac{2}{3} \int d\textbf{\textit{s}}\,N_\mathrm{MP}(\textbf{\textit{s}},\textbf{\textit{b}}),
\end{equation}
where $2/3$ reflects the number of charged particles in the total number of particles.

The jet particle at $\textbf{\textit{J}}_0$ then traverses the medium along the trajectory vector ($\textbf{\textit{l}}$) at an angle $\phi_s$ with respect to the reaction plane before finally colliding with the medium partons at $\textbf{\textit{C}}_0$, thereby transferring a fraction of its momentum. In this process, the number of kicked-partons per jet particle can be formulated as
\begin{equation}\label{eq:Nk}
    n_k(\textbf{\textit{C}}_0,\phi_s,\textbf{\textit{b}})=\int_{0}^{\infty}\frac{dl}{2t}\,\sigma_\mathrm{MP}\,N_\mathrm{MP}(\textbf{\textit{C}}_0,\textbf{\textit{b}}),
\end{equation}
which depends on $N_\mathrm{MP}$ along $\textbf{\textit{l}}$. In addition, since the medium expands in time, we assume that the medium parton density is inversely proportional to the time $t$, which can be approximated as $t\approx t_0+l$. In parallel, the factor of 2 in $t$ comes from the front and back sides in the longitudinal direction. If we then consider the scattering cross-section $\sigma_\mathrm{MP}$ when the jet particles collide with the medium partons, we can finally calculate the $n_k$.

\begin{figure}[t]
    \centering\includegraphics[width=0.38\textwidth]{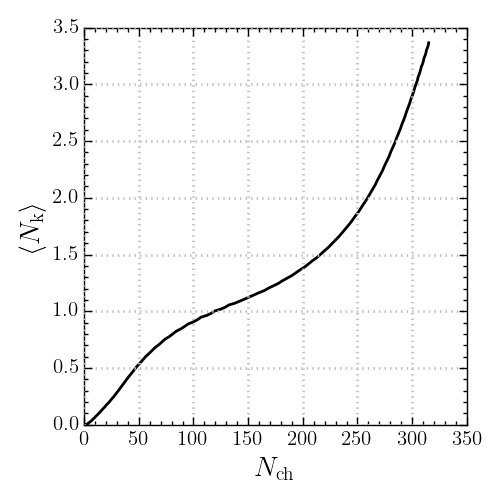}
    \caption{\label{fig:Nch_Nk}Result of the average number of kicked-partons ($\langle N_k \rangle$) as a function of the charged particle multiplicity ($N_\mathrm{ch}$) using the impact parameter.}
\end{figure}

In the real world, the jet particles lose their momentum while passing through the medium not only due to collisions with the medium partons, but also due to the gluon radiation and the absorptive inelastic processes~\cite{Wong:STAR_AuAu_200GeV}. This effect influences the production probability of the jet particles by the factor of $e^{-\zeta n_k(\textbf{\textit{C}}_0,\phi_s,\textbf{\textit{b}})}$, where $\zeta$ is an empirical attenuation coefficient. Consequently, the number of kicked-partons averaged over every $\textbf{\textit{J}}_0$ is given by
\begin{eqnarray}\label{eq:average nk}
    &&N_k(\phi_s,\textbf{\textit{b}}) \nonumber\\
    &&= \frac{\int d\textbf{\textit{J}}_0\,n_k(\textbf{\textit{C}}_0,\phi_s,\textbf{\textit{b}})\,e^{-\zeta n_k(\textbf{\textit{C}}_0,\phi_s,\textbf{\textit{b}})}\,P_\mathrm{jet}(\textbf{\textit{J}}_0,\textbf{\textit{b}})}{\int d\textbf{\textit{J}}_0\,e^{-\zeta n_k(\textbf{\textit{C}}_0,\phi_s,\textbf{\textit{b}})}\,P_\mathrm{jet}(\textbf{\textit{J}}_0,\textbf{\textit{b}})}.
\end{eqnarray}
Lastly, Eq.~\eqref{eq:average nk} is averaged over the angle $\phi_s$ to obtain $\langle N_k \rangle$ as a function of $\textbf{\textit{b}}$:
\begin{equation}\label{eq:final Nk}
    \langle N_k \rangle (\textbf{\textit{b}}) = \frac{1}{\pi/2} \int_{0}^{\pi/2}d\phi_s\,N_k(\phi_s,\textbf{\textit{b}}),
\end{equation}
where the angular range of $\phi_s$ is set from 0 to $\pi/2$ due to the symmetry of the transverse coordinate system.

In conclusion, the MKM has the multiplicity dependence by connecting $N_\mathrm{ch}$ and $\langle N_k \rangle$ through $\textbf{\textit{b}}$ using Eqs.~\eqref{eq:Nch} and \eqref{eq:final Nk}, which is shown in Fig.~\ref{fig:Nch_Nk}. The figure is the result of $\langle N_k \rangle$ as a function of $N_\mathrm{ch}$ and shows the linearly proportional relationship between them.

\begin{table*}[t]
\caption{\label{tab:total parameters}Tabulated physical parameters used in the MKMwM across different experiments and energy scales. The table is divided into two parts: the MKM parameters and the multiplicity (centrality) dependence parameters. The parameters applied to the STAR~\cite{exp:STAR_AuAu_200GeV_1,exp:STAR_AuAu_200GeV_3,exp:STAR_AuAu_200GeV_5,exp:STAR_AuAu_200GeV_6e,exp:STAR_AuAu_200GeV_11} and the PHENIX~\cite{exp:PHENIX_AuAu_200GeV_0,exp:PHENIX_AuAu_200GeV_1} experiments for AuAu collisions at $\sqrt{s_\mathrm{NN}}=0.2$ TeV are taken from Refs.~\cite{Wong:STAR_AuAu_200GeV} and~\cite{Wong:PHENIX_AuAu_200GeV}, respectively. The parameters applied to the CMS~\cite{exp:CMS_pp_7TeV} experiment for $pp$ collisions at $\sqrt{s}=7$ TeV are sourced from Ref.~\cite{Wong:CMS_pp_7TeV}. Our calculated parameters in the CMS experiment for $pp$ collisions at $\sqrt{s}=13$ TeV are listed in the last column.}
\renewcommand{\arraystretch}{1.2}
\begin{ruledtabular}
\begin{tabular}{ccccc}
 & Physical parameter & STAR \& PHENIX & \multicolumn{2}{c}{CMS}\\
 & & AuAu at $\sqrt{s_{NN}}=0.2$ TeV & $pp$ at $\sqrt{s}=7$ TeV & $pp$ at $\sqrt{s}=13$ TeV \\
\colrule
\multirow{5}{*}{MKM} & $q$ & 1.0 \& 0.8 GeV/c & 2.0 GeV/c & 1.2 GeV/c \\
 & $f_R$ & 0.632 & 1.0 & 0.44 $\sim$ 1.89 \\
 & $T$ & 0.50 GeV & 0.70 GeV & 0.77 GeV \\
 & $m_d$ & 1.0 GeV & 1.0 GeV & 1.0 GeV \\
 & $a$ & 0.5 & 0.5 & 0.5 \\
 \colrule
\multirow{5}{*}{\begin{tabular}[c]{@{}c@{}} Multiplicity \\ (Centrality) \\ dependence\end{tabular}} & $R_A\;\&\;R_B$ &  & 0.80 fm & 0.74 fm \\
 & $t_0$ & 0.60 fm/c & 0.43 fm/c & 0.39 fm/c \\
 & $\kappa'$ & 21 & 367 & 236 \\
 & $\sigma_\mathrm{MP}$ & 1.4 mb & 1.4 mb & 1.4 mb \\
 & $\zeta$ & 0.20 & 0.20 & 0.20 \\
\end{tabular}
\end{ruledtabular}
\end{table*}

\begin{figure*}[t]
    \centering
    \includegraphics[width=0.9\textwidth]{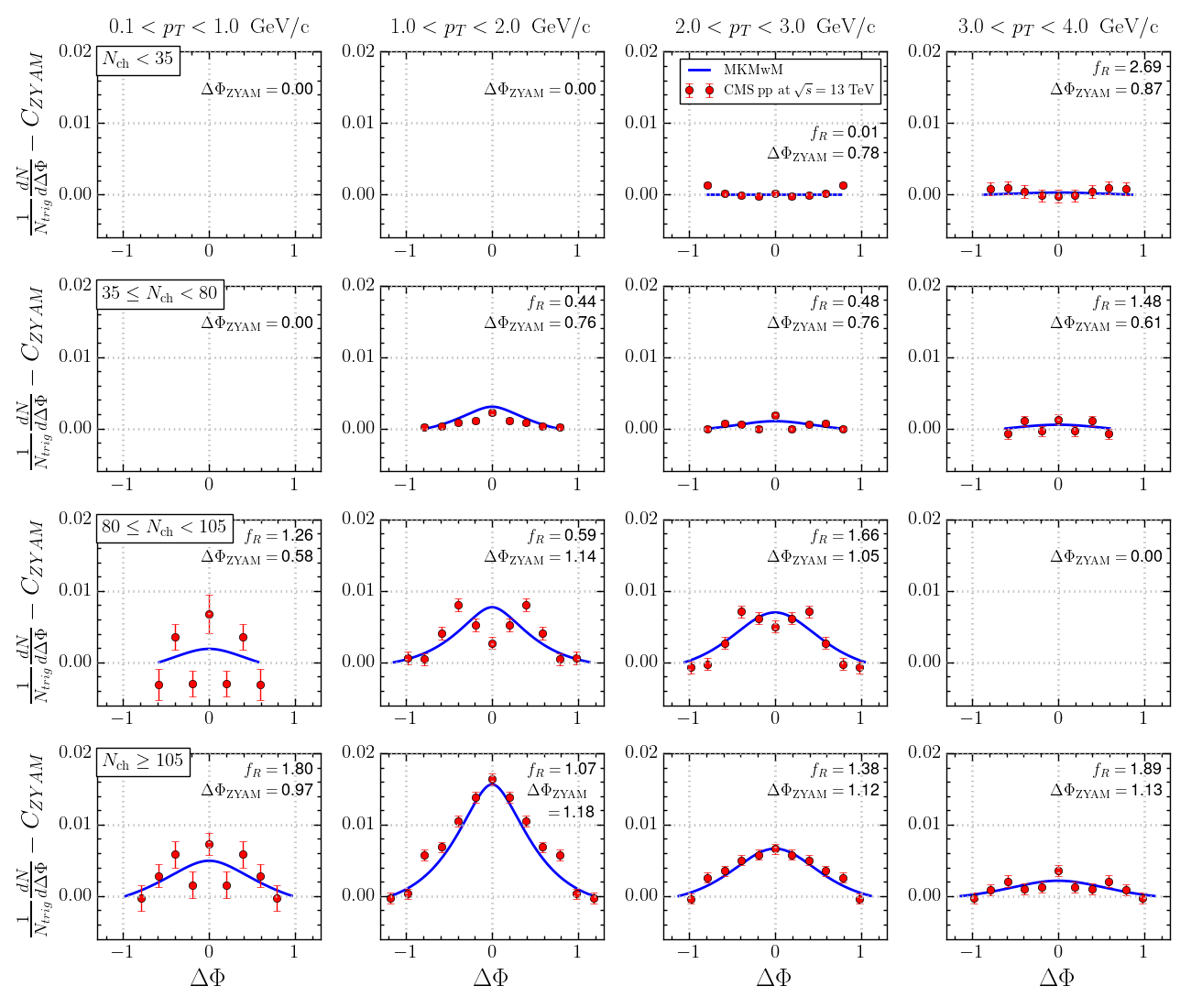}
    \caption{\label{fig:Phi_Correlation_q12}(Color online) Yield per trigger as a function of $\Delta\Phi$, averaged over $2<|\Delta\eta|<4$, in the CMS experiment for $pp$ collisions at $\sqrt{s}=13$ TeV~\cite{exp:CMS_pp_13TeV}, when the $q$ value is 1.2 GeV/c. The configuration consists of each column with four different $p_T$ intervals ($0.1\sim1.0,\;1.0\sim2.0,\;2.0\sim3.0,\;3.0\sim4.0$ in GeV/c) and each row with four different $N_\mathrm{ch}$ ranges ($<35,\;35\sim80,\;80\sim105,\;\ge 105$). The blue curves and the red circles are the theoretical results and the experimental data with uncertainty, respectively. Each panel displays the values of $\Delta\Phi_\mathrm{ZYAM}$ and $f_R$ at the upper right corner. Note that no experimental data is available for bins where $\Delta\Phi_\mathrm{ZYAM}=0$.}
\end{figure*}

\section{ANALYSIS AND PREDICTION}\label{sec:ANALYSIS}
In this section, we apply the MKMwM to the CMS experiment for $pp$ collisions at $\sqrt{s}=13$ TeV~\cite{exp:CMS_pp_13TeV} by introducing some physical extensions. We also discuss our modifications to the MKMwM in comparison with the previous study at $\sqrt{s}=7$ TeV~\cite{Wong:CMS_pp_7TeV} and confirm whether they work properly on the $\Delta\Phi$ distribution for each $p_T$ and $N_\mathrm{ch}$ bins over a wider multiplicity range at higher collision energy. With this foundation, we answer the experimental questions and predict the potential ridge structure at other collision energies in the upcoming Run 3 experiments.

Table.~\ref{tab:total parameters} shows the physical parameters in the MKMwM across different experiments and energy scales and consists of two parts, the MKM parameters and the multiplicity (centrality) dependence parameters. In the third column at the upper and lower section of the table, the MKM and the centrality parameters were obtained from the STAR~\cite{exp:STAR_AuAu_200GeV_1,exp:STAR_AuAu_200GeV_3,exp:STAR_AuAu_200GeV_5,exp:STAR_AuAu_200GeV_6e,exp:STAR_AuAu_200GeV_11} and the PHENIX~\cite{exp:PHENIX_AuAu_200GeV_0,exp:PHENIX_AuAu_200GeV_1} experiments for AuAu collisions at $\sqrt{s_\mathrm{NN}}=0.2$ TeV, as can be seen in Refs.~\cite{Wong:STAR_AuAu_200GeV} and~\cite{Wong:PHENIX_AuAu_200GeV}. Among them, the parameters $q$, $T$, $t_0$, and $\kappa'$ in the fourth column of the table were extended to the LHC collision energy from Ref.~\cite{Wong:CMS_pp_7TeV} for $pp$ collisions at $\sqrt{s}=7$ TeV.

Our parameters in $\sqrt{s}=13$ TeV are listed in the last column of the table, using only $q$ and $f_R$ as free parameters. For the MKM parameters, the values of $q$ and $f_R$ are determined by using the least squares fitting method. The medium temperature at $\sqrt{s}=13$ TeV is adjusted proportionally to the average transverse momentum ($\langle p_T \rangle$) obtained from the fitting function of Ref.~\cite{exp:CMS_pp_7TeV_averpT_2} as in
\begin{equation}\label{eq:medium temperature}
    T=\frac{\langle p_T \rangle_{\sqrt{s}\;=\;13\;\mathrm{TeV}}}{\langle p_T \rangle_{\sqrt{s_{NN}}\;=\;0.2\;\mathrm{TeV}}}\cdot T_{\sqrt{s_{NN}}\;=\;0.2\;\mathrm{TeV}} = 0.77\;\mathrm{GeV},
\end{equation}
where
\begin{equation}\label{eq:average transverse momentum ratio}
    \frac{\langle p_T \rangle_{\sqrt{s}\;=\;13\;\mathrm{TeV}}}{\langle p_T \rangle_{\sqrt{s_{NN}}\;=\;0.2\;\mathrm{TeV}}}=\frac{0.604\;\mathrm{GeV/c}}{0.392\;\mathrm{GeV/c}}=1.54.
\end{equation}
Lastly, we use the same values for $m_d$ and $a$ determined from the STAR and the PHENIX experimental data since they are not sensitive to collision energy.

For the multiplicity dependence parameters, the radii of the colliding proton particles are inferred from the inelastic cross-section, $\sigma_\mathrm{inel}\sim 68.6$ mb, which is a more recent value at $\sqrt{s}=13$ TeV~\cite{exp:CMS_pp_13TeV_cross-sec_3}. Consequently, the combined radius of two colliding protons is given by
\begin{equation}
    R = R_A+R_B = \sqrt{\sigma_\mathrm{inel}/\pi} = 1.48\;\mathrm{fm},
\end{equation}
which leads to estimating the radius of each proton as $R_A=R_B=0.74$ fm. The time $t_0$ required to produce the transverse mass would be inversely proportional to the collision energy because the longitudinal momentum is converted to the transverse momentum more rapidly as the collision energy increases. This effect on $t_0$ at $\sqrt{s}=13$ TeV can be compensated by using the value from Eq.~\eqref{eq:average transverse momentum ratio}, leading to
\begin{equation}\label{eq:time for transverse mass}
    t_{0}=\frac{t_{0,\;\sqrt{s_{NN}}\;=\;0.2\;\mathrm{TeV}}}{1.54}=0.39\;\mathrm{fm/c}.
\end{equation}
The $\kappa'$ value in Eq.~\eqref{eq:kappa'} is reformed by
\begin{equation}\label{eq:kappa simple}
    \kappa'=\frac{\langle N_\mathrm{MP} \rangle}{\langle N_\mathrm{particp} \rangle}=\frac{3}{2} \frac{\langle N_\mathrm{ch} \rangle}{\langle N_\mathrm{particp} \rangle} = 236,
\end{equation}
where $3/2$ is the ratio of total particles to charged particles, and $\langle N_\mathrm{particp} \rangle$ is obtained as 0.5019 from the sum of the thickness function of colliding particles. The $\kappa'$ value is significantly different from that in the previous study since the $\langle N_\mathrm{ch} \rangle$ is determined to be 79 using the most recent accurate results that are given as a fitting function of the collision energy in $pp/\Bar{p}p$ collisions~\cite{average_Nch}. Lastly, we use the same values for $\sigma_\mathrm{MP}$ and $\zeta$ due to their insensitivity to the collision energy at the parton level.

\begin{table}[b]
\caption{\label{tab:Phi matching the average Mul}Multiplicity class for the $\Delta\Phi$ distribution observed in the CMS experiment for $pp$ collisions at $\sqrt{s}=13$ TeV~\cite{exp:CMS_pp_13TeV}. The average number of kicked-partons ($\langle N_k \rangle$) is matched one-to-one with the average charged particle multiplicity ($\langle N_\mathrm{ch} \rangle$).}
\renewcommand{\arraystretch}{1.2}
\begin{ruledtabular}
\begin{tabular}{ccc}
Multiplicity class ($N_\mathrm{ch}$) & $\langle N_\mathrm{ch} \rangle$ & $\langle N_k \rangle$ \\
\colrule
{[2,\;34]}    & 16  & 0.078 \\
{[35,\;79]}   & 58  & 0.561 \\
{[80,\;104]}  & 107  & 1.100 \\
{[105,\;134]} & 131 & 1.309 \\
\end{tabular}
\end{ruledtabular}
\end{table}

\begin{figure*}[t]
    \centering\includegraphics[width=0.7\textwidth]{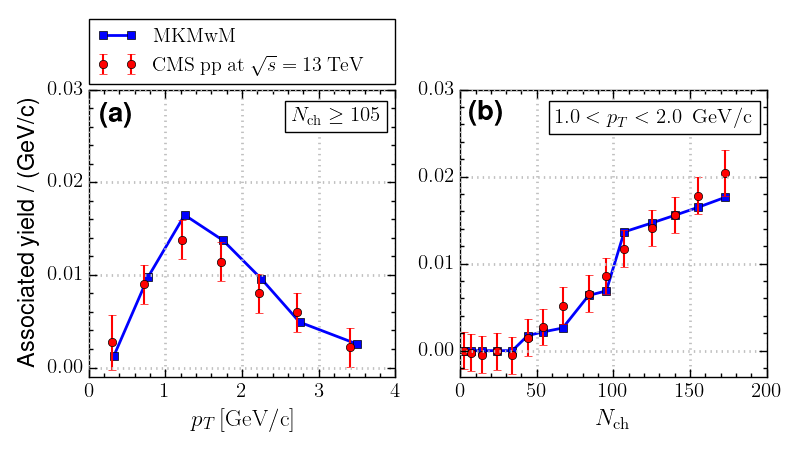}
    \caption{\label{fig:Asso_Yield_q12}(Color online) Associated yield per trigger as a function of $p_T$ and $N_\mathrm{ch}$, averaged over $2<|\Delta\eta|<4$ and integrated over $|\Delta\Phi|<\Delta\Phi_\mathrm{ZYAM}$, in the CMS experiment for $pp$ collisions at $\sqrt{s}=13$ TeV~\cite{exp:CMS_pp_13TeV}, when the $q$ value is 1.2 GeV/c. The blue squares linked by lines are theoretical results, and the red circles are experimental data with uncertainty. (a) The $p_T$ distribution is calculated in the ranges of $N_\mathrm{ch}\ge105$, normalized by the width of the $p_T$ interval. (b) The $N_\mathrm{ch}$ distribution is calculated in the region of $1.0<p_T<2.0$ GeV/c.}
\end{figure*}

The $\Delta\Phi$ distribution averaged over $2<|\Delta\eta|<4$ is depicted in Fig.~\ref{fig:Phi_Correlation_q12}, which presents each column with four different $p_T$ intervals ($0.1\sim1.0,\;1.0\sim2.0,\;2.0\sim3.0,\;3.0\sim4.0$ in GeV/c) and each row with four different $N_\mathrm{ch}$ ranges ($<35,\;35\sim80,\;80\sim105,\;\ge 105$). The blue curves and the red circles are the theoretical results and the experimental data with uncertainty, respectively. As tabulated in the second column of Table.~\ref{tab:Phi matching the average Mul}, the CMS Collaboration has provided $\langle N_\mathrm{ch} \rangle$ for each $N_\mathrm{ch}$ range, weighted by its frequency and corrected for detector effects. In our model analysis, as shown in the last column of the table, we extract $\langle N_k \rangle$ corresponding to these weighted averages from the result between $\langle N_k \rangle$ and $\langle N_\mathrm{ch} \rangle$ using the impact parameter in Fig.~\ref{fig:Nch_Nk}. The CMS collaboration applied the Zero-Yield-At-Minimum (ZYAM) procedure~\cite{exp:CMS_PbPb_276TeV_1,exp:CMS_PbPb_276TeV_2,exp:CMS_PbPb_276TeV_3,exp:CMS_PbPb_276TeV_5,exp:CMS_pp_7TeV,exp:CMS_pp_13TeV,exp:CMS_pp_13TeV_2} to analyze their data. This procedure involves fitting a one-dimensional $\Delta\Phi$ correlation function to a truncated Fourier series extending up to the fifth term, where the minimum of this fitted function is identified as the constant background ($C_\mathrm{ZYAM}$) at $\Delta\Phi=\Delta\Phi_\mathrm{ZYAM}$. The ZYAM procedure determines $\Delta\Phi_\mathrm{ZYAM}$, shown in the upper right corner of each panel in Fig.~\ref{fig:Phi_Correlation_q12}, and sets the ridge yield at $\Delta\Phi_\mathrm{ZYAM}$ to zero by subtracting $C_\mathrm{ZYAM}$ from the one-dimensional $\Delta\Phi$ correlation function. We adopt the same approach in our model calculation by setting the computed ridge yield at $\Delta\Phi_\mathrm{ZYAM}$ to zero, which facilitates direct comparisons between the theoretical results and the experimental data. On the other hand, the $\Delta\Phi_\mathrm{ZYAM}$ depends on experiments, so the survival factor $f_R$ theoretically plays the role of a buffer for the experimental uncertainty in $\Delta\Phi_\mathrm{ZYAM}$ as an overall constant. As a result, the $f_R$ values are determined using the least-squares fitting method, which is applied simultaneously to the experimental data of the $\Delta\Phi$ distribution in Fig.~\ref{fig:Phi_Correlation_q12} and the $p_T$ and $N_\mathrm{ch}$ distributions in Fig.~\ref{fig:Asso_Yield_q12} (a) and (b) altogether. These values, shown in the upper right corner of each panel in Fig.~\ref{fig:Phi_Correlation_q12}, increase with $p_T$ for $p_T>1$ GeV/c, which follows the physical behavior that particles with higher $p_T$ are more likely to survive up to the detector. In contrast, the CMS data exhibit significant fluctuations in the lowest $p_T$ range, which does not ensure a consistent tendency for $f_R$ with $p_T$. Similarly, the ridge yield within $N_\mathrm{ch}<35$ is so small that the resulting $f_R$ values could not have a meaningful interpretation. Moreover, if we focus on the ridge yield in the $N_\mathrm{ch}\ge105$ region because the ridge effect is typically observed in high-multiplicity events, our MKMwM results show good agreement with the experimental data. However, for the $80\le N_\mathrm{ch}<105$ range, the experimental data seem to show two peaks at $p_T>1$ GeV/c, where the MKMwM unfortunately cannot provide a clear explanation for this phenomenon.

For the integrated yield over $|\Delta\Phi|<\Delta\Phi_\mathrm{ZYAM}$, the $p_T$ distribution in the high-multiplicity events, $N_\mathrm{ch}\ge105$, and the $N_\mathrm{ch}$ distribution within $1.0<p_T<2.0$ GeV/c are shown in Fig.~\ref{fig:Asso_Yield_q12} (a) and (b), respectively. The blue squares linked by lines are theoretical results, and the red circles are experimental data with uncertainty. In Fig.~\ref{fig:Asso_Yield_q12} (a), the ridge yield increases in the low $p_T$ region, decreases in the high $p_T$ region, and shows a maximum in the region of $1.0<p_T<2.0$ GeV/c. To further investigate this $p_T$ region, the CMS Collaboration has conducted the $N_\mathrm{ch}$ distribution, where the ridge yield increases linearly with $N_\mathrm{ch}$, as shown in Fig.~\ref{fig:Asso_Yield_q12} (b). Compared to the previous CMS Collaboration study at $\sqrt{s}=7$ TeV, not only is the number of bins for the $p_T$ distribution nearly doubled, but also the $N_\mathrm{ch}$ distribution is about four times finer and covers a wider $N_\mathrm{ch}$ range from (13, 118) to (3, 173). The resulting $q$ value is determined to be 1.2 GeV/c by applying the least-squares fitting method to the experimental data of the $\Delta\Phi$, $p_T$, and $N_\mathrm{ch}$ distributions simultaneously. This $q$ value can provide a theoretical basis for the first question raised by the CMS Collaboration that the ridge yield peaks around $p_T\approx1$ GeV/c and diminishes with increasing $p_T$, which is because the momentum transfer predominantly occurs near 1.2 GeV/c on average. Our results show a satisfactory agreement with these detailed data within the uncertainty by using the parameters listed in Table.~\ref{tab:total parameters}, which leads the MKMwM to be quite successful. Consequently, $\langle N_k \rangle$ in the MKMwM provides a plausible explanation for the second question posed by the CMS Collaboration regarding the ridge yield, which shows an approximately linear increase with $N_\mathrm{ch}$.

\begin{figure*}[t]
    \centering\includegraphics[width=0.9\textwidth]{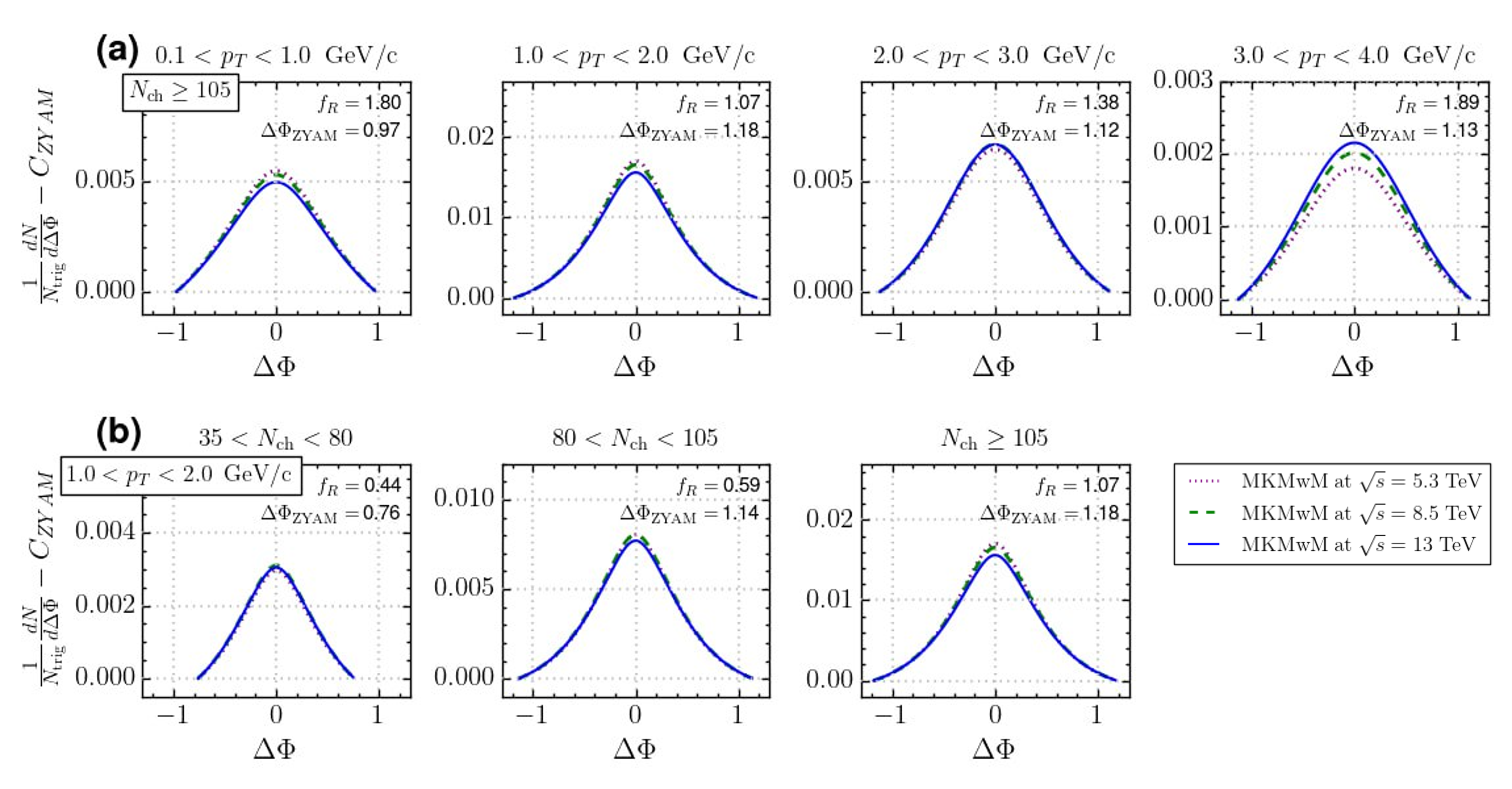}
    \caption{\label{fig:5385_Phi_Correlation_q12}(Color online) Prediction results of the yield per trigger as a function of $\Delta\Phi$, averaged over $2<|\Delta\eta|<4$, in the upcoming LHC Run 3 for $pp$ collisions at $\sqrt{s}=5.3\;\mathrm{and}\;8.5$ TeV, when the $q$ value is 1.2 GeV/c. The purple dotted curves and the green dashed curves are the MKMwM predictions at $\sqrt{s}=5.3$ and $8.5$, respectively, while the blue solid curves are the MKMwM result at $\sqrt{s}=13$ TeV. The same values of $\Delta\Phi_\mathrm{ZYAM}$ and $f_R$ are used as those at $\sqrt{s}=13$ TeV for simple calculation, shown in the upper right corner. (a) The distributions are classified by the $p_T$ ranges ($0.1\sim1.0$, $1.0\sim2.0$, $2.0\sim3.0$, $3.0\sim4.0$ in GeV/c) in the case of $N_\mathrm{ch}\ge105$. (b) The distributions are categorized according to the $N_\mathrm{ch}$ ranges ($35\sim80$, $80\sim105$, $\ge105$) in the region of $1.0<p_T<2.0$ GeV/c. Note that the scale of each figure is different.}
\end{figure*}

\begin{figure*}[t]
    \centering\includegraphics[width=0.7\textwidth]{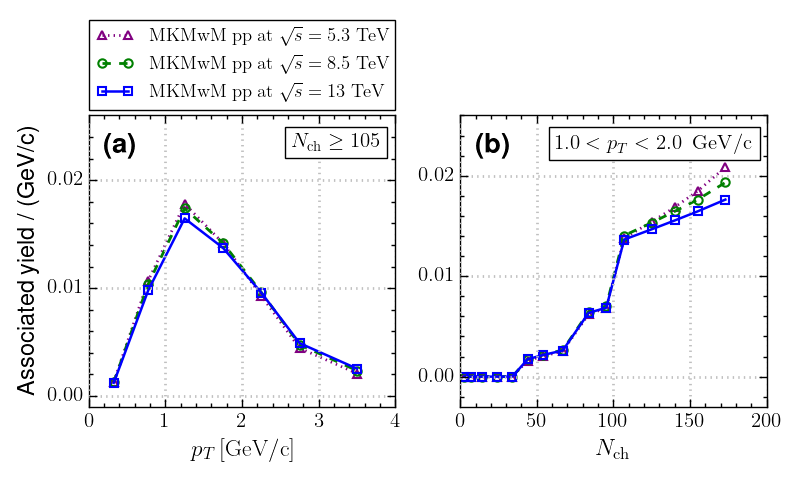}
    \caption{\label{fig:5385_Asso_Yield_q12}(Color online) Predicted results of the associated yield per trigger as a function of $p_T$ and $N_\mathrm{ch}$, averaged over $2<|\Delta\eta|<4$ and integrated over $|\Delta\Phi|<\Delta\Phi_\mathrm{ZYAM}$, in the upcoming LHC Run 3 for $pp$ collisions at $\sqrt{s}=5.3$ and $8.5$ TeV, when the $q$ value is 1.2 GeV/c. The purple triangles linked by dotted lines and the green circles linked by dashed lines represent the MKMwM predictions at $\sqrt{s}=5.3$ and $8.5$ TeV, respectively, while the blue squares linked by solid lines signify the MKMwM result at $\sqrt{s}=13$ TeV. (a) The $p_T$ distribution is presented in the high-multiplicity events, $N_\mathrm{ch}\ge105$. (b) The $N_\mathrm{ch}$ distribution is shown in the region of $1.0<p_T<2.0$ GeV/c.}
\end{figure*}

In addition, the CMS Collaboration compared the ridge yield for $pp$ collisions at $\sqrt{s}=13$ TeV with that at $\sqrt{s}=7$ TeV and found considerable similarity between them~\cite{exp:CMS_pp_13TeV}. Consequently, they suggested that the ridge structure for $pp$ collisions shows no significant dependence on the collision energy. Within the framework of the MKMwM, this suggestion is interpreted to mean that the $q$ value is not susceptible to variations in the collision energy. To confirm it, we calculated the $q$ value at $\sqrt{s}=7$ TeV using the least-squares fitting method, and the result was 1.1 GeV/c, which is close to the value obtained at $\sqrt{s}=13$ TeV. This consistency supports our premise that the $q$ value would be steady across different collision energies.

However, our $q$ value of 1.2 GeV/c, at $\sqrt{s}=13$ TeV deviates considerably from 2.0 GeV/c proposed by the previous study at $\sqrt{s}=7$ TeV~\cite{Wong:CMS_pp_7TeV}. This discrepancy can be primarily attributed to three key factors:
\begin{enumerate}
    \item We calculate the $N_\mathrm{ch}$ distribution both from the $p_T$ and $\Delta\Phi$ dependence. On the other hand, the previous study directly calculates this distribution only from the $p_T$ dependence. These extra resources enable a more precise determination of the $q$ value.
    \item It is noteworthy that more recent experimental findings regarding $\langle N_\mathrm{ch} \rangle$ facilitate comparatively accurate estimation of the $\kappa'$ than the one employed by the previous study. This enhanced precision makes the $q$ value more plausible.
    \item While the previous study assumes $f_R$ as a constant, we treat it as a free parameter depending on $p_T$ and $N_\mathrm{ch}$, which not only gives us room for uncertainty coming from an experimental analysis such as the ZYAM procedure but also proposes that $f_R$ has a $p_T$ dependence.
    \item For the thickness function for the proton, the previous study uses a sharp-cutoff shape~\cite{Wong:CMS_pp_7TeV} for a nucleus, but we use the proton charge density, which is more compatible for the $pp$ collisions.
\end{enumerate}
Considering these differences, our $q$ value of 1.2 GeV/c appears to be reasonable and well-founded.

Currently, the LHC Run 3 is conducting measurements of PbPb collisions at $\sqrt{s_{NN}}=5.3$ TeV, $p$Pb collisions at $\sqrt{s_{NN}}=8.5$ TeV, and $pp$ collisions at the same center-of-mass energies as the PbPb and $p$Pb reference samples~\cite{ALICE_Run3}. With these ongoing experiments, it is possible to make theoretical predictions about the long-range near-side ridge structures for $pp$ collisions at $\sqrt{s}=5.3$ and $8.5$ TeV. Specifically, we expect the $q$ value to maintain an approximately constant value of 1.2 GeV/c, regardless of the variations in collision energies, which would be beneficial to minimize the degrees of freedom in the parameters. In the prediction process, $T$ increases according to the collision energy, and $t_0$ is inversely proportional to the collision energy, as mentioned in Eqs.~\eqref{eq:medium temperature} and \eqref{eq:time for transverse mass}, respectively.

The predicted results for the $\Delta\Phi$ distribution are exhibited in Fig.~\ref{fig:5385_Phi_Correlation_q12}, where the ranges are grouped in the same way as at $\sqrt{s}=13$ TeV. The purple dotted lines and the green dashed lines represent the MKMwM predictions at $\sqrt{s}=5.3$ and $8.5$ TeV, respectively, while the blue solid lines show the MKMwM results at $\sqrt{s}=13$ TeV for comparison. Although we use the same values of $\Delta\Phi_\mathrm{ZYAM}$ and $f_R$ as those at $\sqrt{s}=13$ TeV in this prediction by assuming the equivalent condition, it should be noted that the $\Delta\Phi_\mathrm{ZYAM}$ may be subject to arbitrary adjustments, and the $f_R$ values may vary according to experiments. In addition, the scale of each figure is different for easy comparison. As depicted in Fig.~\ref{fig:5385_Phi_Correlation_q12} (a), the yield at lower collision energy slightly exceeds that at higher collision energy, which is because the larger proportion of low $p_T$ particles is relatively prevalent at lower collision energies, such that yields at $\sqrt{s}=5.3$, 8.5, and 13 TeV are in order. Furthermore, the trend is reversed at $p_T>3.0$ GeV/c because, at the higher collision energies, there is a relatively larger fraction of high $p_T$ particles. It is also important to note that the yield is much bigger at $1.0<p_T<2.0$ GeV/c than that at other ranges, which is the direct result from $q=1.2$ GeV/c. On the other hand, in Fig.~\ref{fig:5385_Phi_Correlation_q12} (b), the yield increases with $N_\mathrm{ch}$, which makes it clear that $\langle N_k \rangle$ is proportional to $N_\mathrm{ch}$.

A more detailed examination of this trend is possible in the $p_T$ and the $N_\mathrm{ch}$ distributions, as depicted in Fig.~\ref{fig:5385_Asso_Yield_q12}, where the purple triangles linked by dotted lines and the green circles linked by dashed lines represent the MKMwM predictions at $\sqrt{s}=5.3$ and $8.5$ TeV, respectively, while the blue squares linked by solid lines indicate the MKMwM results at $\sqrt{s}=13$ TeV. For the $p_T$ distribution in Fig.~\ref{fig:5385_Asso_Yield_q12} (a), it is clear that the difference in the ridge yield between collision energies is most substantial in the range of $1.0<p_T<2.0$ GeV/c. For the $N_\mathrm{ch}$ distribution in Fig.~\ref{fig:5385_Asso_Yield_q12} (b), the difference is especially pronounced in high-multiplicity events.

\section{CONCLUSIONS AND DISCUSSIONS}\label{sec:CONCLUSIONS AND DISCUSSIONS}
Previously, the long-range near-side ridge structure in heavy-ion collisions was explained by the hydrodynamic models driven by the QGP. However, since the LHC era, this structure has also appeared in small systems that are not enough to produce the QGP matter, which is difficult to describe by the hydrodynamic models. Consequently, a variety of models are raised to give a theoretical basis for this structure, but there is no consensus on it. Naturally, the Momentum Kick Model (MKM) has well explained this structure in the STAR~\cite{exp:STAR_AuAu_200GeV_1,exp:STAR_AuAu_200GeV_3,exp:STAR_AuAu_200GeV_5,exp:STAR_AuAu_200GeV_6e,exp:STAR_AuAu_200GeV_11}, the PHENIX~\cite{exp:PHENIX_AuAu_200GeV_0,exp:PHENIX_AuAu_200GeV_1}, the PHOBOS~\cite{exp:PHOBOS_AuAu_200GeV_1}, the CMS~\cite{exp:CMS_PbPb_276TeV_1,exp:CMS_pp_13TeV}, and the ATLAS~\cite{exp:ATLAS_pp_276TeV} experimental data simply by a kinematic process of collisions between jet particles and medium partons. The MKM also required a multiplicity dependence because the ridge structure is prominent for high multiplicity events. Therefore, the previous study applied the MKM with multiplicity (MKMwM) to the CMS experiment for $pp$ collisions at $\sqrt{s}=7$ TeV, which showed good agreement with the data. However, it needs to be validated to higher collision energies and a wider range of multiplicity with finer data points, and its application to the CMS experiment for $pp$ collisions at $\sqrt{s}=13$ TeV is reported in this paper.

This study differs from the previous one at $\sqrt{s}=7$ TeV by its experimental and theoretical aspects. Experimentally, the CMS Collaboration performed not only the $p_T$ and $N_\mathrm{ch}$ distributions but also the $\Delta\Phi$ distribution and analyzed their data with finer bins, which are consistent with our goal of extending the validity of the MKMwM. Theoretically, we find the optimal values of $q$ and $f_R$ using the least-squares fitting method and realistically tune the values of $R$ and $\kappa'$ using the latest experimental data. Furthermore, we show quantitatively that the $f_R$ has a $p_T$ dependence, which can be achieved in future analyses. The application of this study is quite successful and enhances the validity of the MKMwM by providing theoretical answers to the two questions posed by the CMS Collaboration. Moreover, by fixing the most sensitive $q$ value as 1.2 GeV/c through the CMS conjecture in $pp$ collisions, we focus our predictive analysis on two collision energies currently being conducted in the LHC Run 3 experiment. Although using the same values of $\Delta\Phi_\mathrm{ZYAM}$ and $f_R$ as those at $\sqrt{s}=13$ TeV may introduce some uncertainties, we believe that future experimental analyses will reveal the characteristics of the ridge behavior consistent with our predictions.

While the MKMwM provides a plausible mechanism for the ridge phenomenon observed in $pp$ collisions at LHC energies, it remains uncertain whether this mechanism is equally applicable to heavy-ion collisions at the same energy level. Due to the notable differences between heavy-ion collisions and $pp$ or $p$Pb collisions, the parameters including the system size, medium temperature, and the number of charged particles may need to be extended to describe the ridge structure in heavy-ion collisions. Therefore, the feasibility of the MKMwM mechanism as an explanatory model for the ridge phenomenon in heavy-ion collisions needs to be further investigated.

\begin{acknowledgments}
This work is supported by the National Research Foundation of Korea (NRF) grant from the Ministry of Science and ICT (MSIT) of the Government of Korea (No. NRF-2008-00458) and the Inha University Research Fund. The authors appreciate Dr. Wong for his valuable discussion.
\end{acknowledgments}


\bibliography{main}

\end{document}